\documentclass[twocolumn,amssymb, nobibnotes, showpacs, superscriptaddress, aps, prd]{revtex4}
\usepackage{amsmath}
\usepackage{epsfig}

\usepackage{graphicx}% Include figure files
\usepackage{dcolumn}% Align table columns on decimal point
\usepackage{bm}% bold math
\usepackage{color}
\usepackage[bookmarks=true,
   colorlinks=true,
   linkcolor=blue,
   urlcolor=blue,
   citecolor=blue,
   bookmarks=true,
   hyperindex=true
]{hyperref}

\begin{document}

\title{Competition effects of multiple quantum paths in an atom interferometer}
\author{Si-Bin Lu}
\affiliation{State Key Laboratory of Magnetic Resonance and Atomic and Molecular Physics, Wuhan Institute of Physics and Mathematics, Chinese Academy of Sciences, Wuhan 430071, China}
\affiliation{Center for Cold Atom Physics, Chinese Academy of Sciences, Wuhan 430071, China}
\affiliation{School of Physics, University of Chinese Academy of Sciences, Beijing 100049, China}

\author{Zhan-Wei Yao}
\affiliation{State Key Laboratory of Magnetic Resonance and Atomic and Molecular Physics, Wuhan Institute of Physics and Mathematics, Chinese Academy of Sciences, Wuhan 430071, China}
\affiliation{Center for Cold Atom Physics, Chinese Academy of Sciences, Wuhan 430071, China}

\author{Run-Bing Li}
\email[]{rbli@wipm.ac.cn}
\affiliation{State Key Laboratory of Magnetic Resonance and Atomic and Molecular Physics,
Wuhan Institute of Physics and Mathematics, Chinese Academy of Sciences, Wuhan
430071, China}
\affiliation{Center for Cold Atom Physics, Chinese Academy of Sciences, Wuhan 430071, China}

\author{Jun Luo}
\affiliation{State Key Laboratory of Magnetic Resonance and Atomic and Molecular Physics, Wuhan Institute of Physics and Mathematics, Chinese Academy of Sciences, Wuhan 430071, China}
\affiliation{Center for Cold Atom Physics, Chinese Academy of Sciences, Wuhan 430071, China}

\author{Sachin Barthwal}
\affiliation{State Key Laboratory of Magnetic Resonance and Atomic and Molecular Physics, Wuhan Institute of Physics and Mathematics, Chinese Academy of Sciences, Wuhan 430071, China}
\affiliation{Center for Cold Atom Physics, Chinese Academy of Sciences, Wuhan 430071, China}
\affiliation{School of Physics, University of Chinese Academy of Sciences, Beijing 100049, China}

\author{Hong-Hui Chen}
\affiliation{State Key Laboratory of Magnetic Resonance and Atomic and Molecular Physics, Wuhan Institute of Physics and Mathematics, Chinese Academy of Sciences, Wuhan 430071, China}
\affiliation{Center for Cold Atom Physics, Chinese Academy of Sciences, Wuhan 430071, China}
\affiliation{School of Physics, University of Chinese Academy of Sciences, Beijing 100049, China}

\author{Ze-Xi Lu}
\affiliation{State Key Laboratory of Magnetic Resonance and Atomic and Molecular Physics, Wuhan Institute of Physics and Mathematics, Chinese Academy of Sciences, Wuhan 430071, China}
\affiliation{Center for Cold Atom Physics, Chinese Academy of Sciences, Wuhan 430071, China}
\affiliation{School of Physics, University of Chinese Academy of Sciences, Beijing 100049, China}

\author{Jin Wang}
\email[]{wangjin@wipm.ac.cn}
\affiliation{State Key Laboratory of Magnetic Resonance and Atomic and Molecular Physics,
Wuhan Institute of Physics and Mathematics, Chinese Academy of Sciences, Wuhan
430071, China}
\affiliation{Center for Cold Atom Physics, Chinese Academy of Sciences, Wuhan 430071, China}

\author{Ming-Sheng Zhan}
\email[]{mszhan@wipm.ac.cn}
\affiliation{State Key Laboratory of Magnetic Resonance and Atomic and Molecular Physics,
Wuhan Institute of Physics and Mathematics, Chinese Academy of Sciences, Wuhan
430071, China}
\affiliation{Center for Cold Atom Physics, Chinese Academy of Sciences, Wuhan 430071, China}

\date{\today}

\begin{abstract}
  We present an observation of competition effect among multiple quantum paths in a Raman-type Mach-Zehnder atom interferometer. By measuring the contrast of interference fringes, the competition effect among multiple interference paths is experimentally investigated. Due to the phase competition, the contrast periodically oscillates when modulating either the phase or the interrogation time between Raman pulses. The multiple quantum paths form because of the imperfect population transfer efficiency in stimulated Raman transitions, and are verified by modulating the duration of Raman pulses. The contrast could be optimized by suppressing the phase competition.
\end{abstract}

\pacs{37.25.+k, 03.75.Dg, 32.80.Qk}
%\keywords{}

\maketitle

\section{Introduction}

The atom interferometer \cite{Kasevich1991a} is an effective measuring tool in many scientific and technical fields. It has been used to test the weak equivalence principle \cite{Kasevich2002a,Zhou2015a}, measure the atomic polarizability \cite{Ekstrom1995a}, probe quantum and thermal noise of the interacting many-body system \cite{Hofferberth2008a}, and study relaxation and prethermalization of the isolated quantum system \cite{Gring2012a}. A three-Raman-pulse sequence is usually applied to build a Mach-Zehnder (M-Z) atom interferometer which can cancel out the inhomogeneous dephasing and unwanted clock effects \cite{Hahn1950a,Andersen2003a}. The M-Z interferometer enables the high sensitivity for measurement of inertial forces. Further applications include measuring the Earth's gravity \cite{Peters1999a,Hu2013a,Freier2016a,Gillot2014a,Sorrentino2014a,Biedermann2015a} and rotation \cite{Canuel2006a,Stockton2011a,Dutta2016a,Yao2016a,Berg2015a,Gustavson2000a,Yao2018a}. To improve the sensitivity, advanced Raman-pulse sequences were developed. In rotation measurements, a four-Raman-pulse interferometer was used to cancel the gravity effect \cite{Stockton2011a,Dutta2016a}, and a composite-Raman-pulse sequence was applied to enlarge the interference area \cite{Berg2015a}. Multi-pulse interferometers were also studied \cite{Weitz1996a,Hinderth1998a,Aoki2001a,Vincent2010a}.

\vskip 5pt
\noindent

In an atom interferometer, the Raman pulse plays a key role as a beam splitter or mirror. However, due to the spatial intensity distribution of laser fields, the population transfer efficiency in stimulated Raman transitions is less than $90$ $\%$ \cite{Du2016a}. Because of the low transfer efficiency, it is very difficult to generate an ideal Raman pulse, which makes multiple interference paths formed after a Raman-pulse sequence is applied. For example, the imperfect Raman-pulse sequence causes multiple interference paths in the double diffraction scheme \cite{Malossi2010a} and multiple interference loops in a four-Raman-pulse interferometer \cite{Stockton2011a,Takase2008a}. As more Raman pulses such as composite Raman pulses are applied \cite{Berg2015a}, only a part of the atoms participate in the main interference path and more atoms join in unwanted interference paths. These multi-path interferometers exhibit different phase sensitivity to rotations and accelerations, which can lead to systematic errors in absolute measurements \cite{Takase2008a}. Thus, unwanted paths in typical cases are avoided by using the blow-away light or by changing the interrogation time \cite{Malossi2010a,Berg2015a,Stockton2011a}. However, as more pulses are applied, more unwanted paths form besides the main interferometer. In a multi-pulse interferometer, it is difficult to distinguish unwanted paths from the main interferometer. Multiple interference paths cause the phase competition among them, which affects the contrast of interference fringes.

\vskip 5pt
\noindent

In this paper, we design a method to quantitatively investigate the competition effect among multiple quantum paths. For simplicity, the experiment is implemented in a M-Z interferometer. The interference fringes are observed when the atoms are manipulated by the co-propagating Raman beams. By modulating either the phase or the interrogation time, we observe that the contrasts of interference fringes vary periodically. We carefully analyze this phenomenon theoretically as well as experimentally. The competition of multiple quantum paths is verified by modulating the duration of Raman pulses. We confirmed that, if the population transfer efficiency in stimulated Raman transitions is imperfect, the multiple quantum paths form and the fringe contrast is influenced by the phase competition among them.
\section{Experimental setup}

\begin{figure}[htp]
	\centering
		\includegraphics[width=0.42\textwidth]{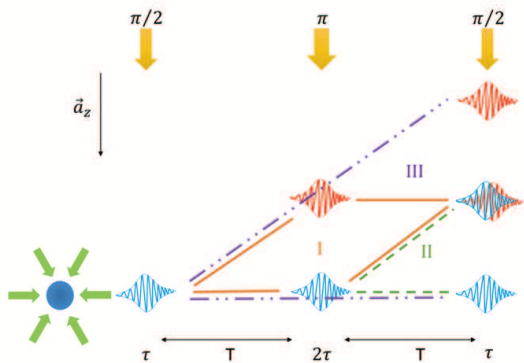}
	\caption{Schematic diagram of multiple interference paths. Three interference paths ($\mathrm{I}$, $\mathrm{II}$ and $\mathrm{III}$) form due to the low transfer efficiency of Raman transitions when the Raman-pulse sequence ($\pi/2-\pi-\pi/2$) is applied. There is the phase competition for the different interrogation time ($T$), duration of Raman pulses ($2\tau$) and phase of Raman lasers ($\phi_{i}$). }%
	\label{fig-1}%
\end{figure}

The schematic diagram is shown in Fig.\textit{\ref{fig-1}}. The $^{87}$Rb atoms are trapped in a magneto-optical trap (MOT), and are cooled to $6$ $\mu$K by polarization gradient cooling. After turning off the cooling lights and the trapping magnetic field, atoms freely fall along the gravity direction. Three pairs of Helmholtz coils are used to compensate for the residual magnetic field \cite{Li2008a}. The experiment is performed in a Raman-type M-Z interferometer. The atoms are prepared in the ground state $|F=1,m=0\rangle$. A pair of Raman lasers with the polarization configuration ($\sigma^{+},\sigma^{+}$) are co-propagating along the direction of gravity. Thus loss of the contrast caused by the gravity and its gradient can be ignored \cite{Albert2014a,Amico2017a}. The Raman lasers are locked by an optical phase-locked loop(OPLL) \cite{Marino2008a}. They are combined together with a polarization beam splitter, and sent to the atom interference area with a polarization maintaining fiber. The Raman lasers couple two ground states $|F=1,m=0\rangle$ and $|F=2,m=0\rangle$ with the common excited state $|F'=1,m'=1\rangle$. The one-photon detuning ($\vartriangle$) is $1.4$ GHz, and the two-photon detuning ($\delta$) is adjusted by the reference frequency of the OPLL. After the Raman pulse sequence ($\pi/2-\pi-\pi/2$) is applied, fringes are observed by measuring the population in the state $|F=2,m=0\rangle$ with laser-induced-fluorescence signals.

\vskip 5pt
\noindent

When the atoms are manipulated by a Raman pulse with duration $2\tau$, the probability of population transfer is $P=\frac{\Omega^{2}_{R}}{\Omega^{2}_{R}+\delta^{2}}\sin^{2}(\sqrt{\Omega^{2}_{R}+\delta^{2}}\tau)$, where $\Omega_{R}$ is the effective Rabi frequency. The population periodically varies with the interaction time, and it can be well transferred in principle when the two-photon resonance is matched. However, it is difficult to completely transfer the population due to the spatial intensity distribution of laser fields. In our experiment, the transfer efficiency ($\eta$) is $(82\pm0.2)\%$ in stimulated Raman transitions, which is measured by simultaneously detecting the population both in $|F=1,m=0\rangle$ and in $|F=2,m=0\rangle$ \cite{ref1}. When the atoms are split into a coherence superposition of the spin states by the first $\pi/2$ pulse, about $18$ $\%$ atoms still stay in the initial state. Due to the imperfect transfer efficiency, after the atoms in both of the superposition and initial states are manipulated by the middle $\pi$ pulse, they stay in the different quantum paths. When the second $\pi/2$ pulse is applied, the multiple interference paths form. From Fig.\textit{\ref{fig-1}}, there are three typical interference paths ($\mathrm{I}$, $\mathrm{II}$ and $\mathrm{III}$). The M-Z interference path ($\mathrm{I}$, orange solid lines) is built by the first $\pi/2$, middle $\pi$ and second $\pi/2$ pulses. One Ramsey interference path ($\mathrm{II}$, green dashed lines) forms by the middle $\pi$ and second $\pi/2$ pulses, and another Ramsey interference path ($\mathrm{III}$, purple dot-dashed lines) forms by the first $\pi/2$ and second $\pi/2$ pulses. These three paths contribute to the interferometer output, and cause the interference competition among them. The competition effect is observed and further investigated by measuring the contrast.

\section{results and discussion}

The competition effect is investigated by measuring the dependence of the contrast on the relative phases among the three interference paths in the M-Z atom interferometer.  We use a new formula to describe the interference of multiple quantum paths mentioned above. For the multiple interference paths as shown in Fig.\textit{\ref{fig-1}}, the total output signal is contributed by one M-Z interferometer and two Ramsey interferometers. The atoms in the paths $\mathrm{I}$ and $\mathrm{III}$ are coherent, but they are incoherent with the atoms in the path $\mathrm{II}$ \cite{Erik2000a,Ghosh2006a,Benatti2002a}. When finite length of Raman pulses is considered by the sensitivity function \cite{Patrick2008a}, the population in the state $|F=2,m=0\rangle$ is written as
\begin{eqnarray}\label{eq1}
P &\!=\!& \frac{1}{2}\!+\!e^{-\gamma_{1}t}\frac{A_{1}}{2}\cos(\phi_{1}\!-\!2\phi_{2}\!+\!\phi_{3})\nonumber\\
& &\!-\!e^{-\gamma_{2}t}\frac{A_{2}}{2}\cos[\phi_{2}\!-\!\phi_{3}\!+\!\delta(T\!+\!\frac{4\tau}{\pi})]\nonumber\\
& &\!-\!e^{-\gamma_{3}t}\frac{A_{3}}{2}\cos[\phi_{1}\!-\!\phi_{3}\!+\!\delta(2T\!+\!2\tau\!+\!\frac{4\tau}{\pi})]\!+\!B,
\end{eqnarray}
where, $A_i$ $(i=1, 2, 3)$ are the amplitudes of three interferometers ($\mathrm{I}$, $\mathrm{II}$ and $\mathrm{III}$), which depend on the population transfer efficiency in Raman transitions. Here, $A_1=\eta^2$, $A_2=2\sqrt{\eta}(1-\eta)^\frac{3}{2}$ and $A_3=\eta(1-\eta)$. $B=\frac{1}{2}(1-A_1-A_2-A_3)$ is the background offset. $\gamma_{i}$ $(i=1,2,3)$ are the decoherent coefficients of the three interference paths, $\phi_{i}$ $(i=1, 2, 3)$ are the phases of three Raman pulses, and $T$ is the interrogation time between two consecutive pulses. The competition among the three interferometers is mainly manifested in the contrast of interference fringes. For $\delta\neq0$, the relative phases can be modulated by the phase and interrogation time of Raman pulses. When $\phi_2$ is modulated, the Ramsey interferometer ($\mathrm{III}$) is not affected, and the competition between the M-Z interferometer ($\mathrm{I}$) and the Ramsey interferometer ($\mathrm{II}$) can be observed. When $T$ is modulated, there is the competition among multiple interference paths.

\vskip 5pt
\noindent

\begin{figure}[htp]
	\centering
		\includegraphics[width=0.46\textwidth]{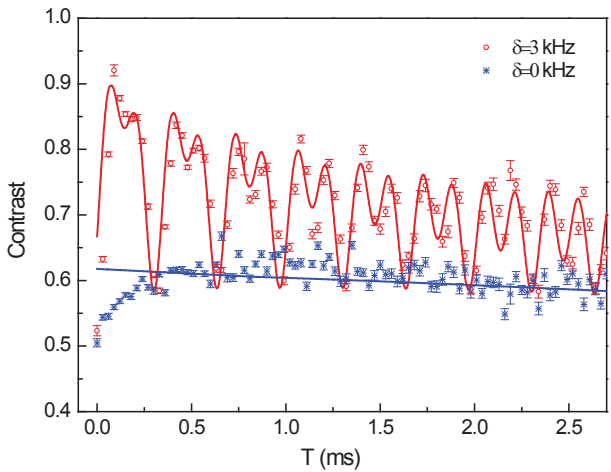}
	\caption{Interference competition for different two-photon detuning at $\delta=0$ $kHz$ (blue stars) and $\delta=3$ $kHz$ (red circles). The contrast shows two oscillation periods for $\delta=3$ $kHz$.}%
	\label{fig-2}%
\end{figure}

\vskip 5pt
\noindent

From Eq.(\ref{eq1}), the contrast of the interference fringe is influenced by the relative phases among the multiple interference paths. Especially, it oscillates with the interrogation time for the non-zero two-photon detuning. We compare the contrast of the M-Z interferometer with and without two-photon detuning. In the experiment, the durations of the $\pi/2$ and $\pi$ pulses are $30$ $\mu$s and $60$ $\mu$s, respectively. When the interrogation time is fixed, after the atoms are manipulated by the three-Raman-pulse sequence ($\pi/2-\pi-\pi/2$), the fringes are observed by scanning the phase of the third pulse at $\delta=0$ kHz and $\delta=3$ kHz, respectively. The contrasts are obtained by fitting the interference fringes with sine functions, and the error bars given by the fitted errors \cite{ref2}. The contrasts for the different interrogation times are shown in Fig.\textit{\ref{fig-2}}. For $\delta$ = $3$ kHz, the contrast of the M-Z interferometer oscillates with two periods of $167$ $\mu$s and $334$ $\mu$s due to phase competitions from two Ramsey interferometers (red circles), which can be explained by Eq.(\ref{eq1}). Two periodical oscillations are caused by the phase variations of two Ramsey interferometers as the interrogation time increased. When the interrogation time is $143$ $\mu$s, the M-Z interferometer ($\mathrm{I}$) is mainly competing with the Ramsey interferometer ($\mathrm{III}$) because of their opposite phase \cite{li2016a}. When the interrogation time is $310$ $\mu$s, the M-Z interferometer ($\mathrm{I}$) is competing with both of Ramsey interferometers ($\mathrm{II}$ and $\mathrm{III}$), which causes a large reduction of the contrast. As the interrogation time increases, two periodic oscillations are more distinct because the competition increases from the interference path $\mathrm{III}$. For $\delta=0$ kHz, the contrast does not oscillate because the relative phases are fixed among three interference paths (blue stars). The decoherence coefficient in the M-Z interferometer ($\gamma_{1}$) is smaller than those in two Ramsey interferometers ($\gamma_{2}$ and $\gamma_{3}$), which causes a slow reduction of the contrast as the interrogation time increased. The solid lines are the contrasts calculated by $C=(P_{max}-P_{min})/(P_{max}+P_{min})$ with $\eta=82\%$ according to Eq.(\ref{eq1}), after $\gamma_{i}$ is obtained by fitting the experimental data \cite{Du2016a}.

\vskip 5pt
\noindent

\begin{figure}[htp]
	\centering
		\includegraphics[width=0.8\textwidth]{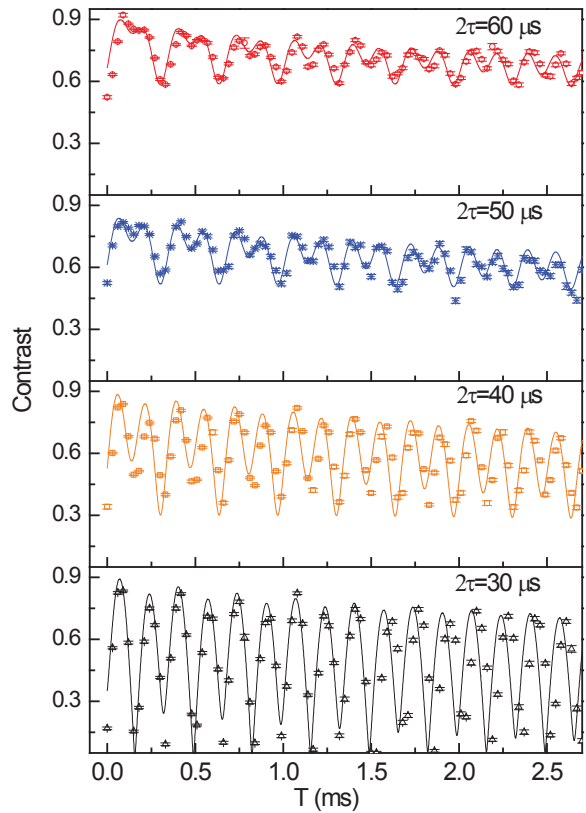}
     	\caption{ Interference competition for different Raman durations at $\delta=3$ kHz. The contrast oscillates as the interrogation time for $2\tau=60$ $\mu$s (red circles), $50$ $\mu$s(blue stars), $40$ $\mu$s(orange squares) and $30$ $\mu$s(black triangles). }%
	\label{fig-3}%
\end{figure}

\vskip 5pt
\noindent

To further study the competition among the three interference paths, the contrast for different durations of the middle Raman pulse is measured. For $\delta=3$ kHz, the contrast oscillates as shown in Fig.\textit{\ref{fig-3}}. Here, the durations of two $\pi/2$ pulses are both $30$ $\mu$s, and the durations of the middle Raman pulse are $60$ $\mu$s (red circles), $50$ $\mu$s (blue stars), $40$ $\mu$s (orange squares) and $30$ $\mu$s (black triangles), respectively. For $2\tau=60$ $\mu$s, the contrast for the M-Z interferometer is maximal because of the minimum contribution from the interference paths ($\mathrm{II}$ and $\mathrm{III}$). In fact, the contrast still oscillates due to the competition when considering the population transfer efficiency of Raman transitions. The competitive effect gradually increases when the duration of the middle Raman pulse is changed from $2\tau=60$ to $30$ $\mu$s. The oscillation periods and amplitudes depend on the competition between the M-Z ($\mathrm{I}$) and Ramsey ($\mathrm{II}$) interferometers, and that between the M-Z ($\mathrm{I}$) and Ramsey ($\mathrm{III}$) interferometers. For $2\tau=30$ $\mu$s, the contrast strongly oscillates due to the maximum competitions from two Ramsey interferometers ($\mathrm{II}$ and $\mathrm{III}$), because more atoms participate in the interference path $\mathrm{III}$ and a balanced Ramsey interferometer is formed in the path $\mathrm{II}$. The phase competitions among three interference paths are verified by modulating the duration of the middle pulse. Here, the population transfer efficiencies are simulated by changing the duration of the middle pulse. According to Eq.(\ref{eq1}) and the experimental parameters, the solid lines are fitted results with the same method as in Fig.\textit{\ref{fig-2}}. In an atom interferometer, the contrast is influenced by the imperfect population transfer efficiency and durations of Raman pulses due to the phase competitions among multiple interference paths.

\vskip 5pt
\noindent

\begin{figure}[htp]
	\centering
		\includegraphics[width=0.46\textwidth]{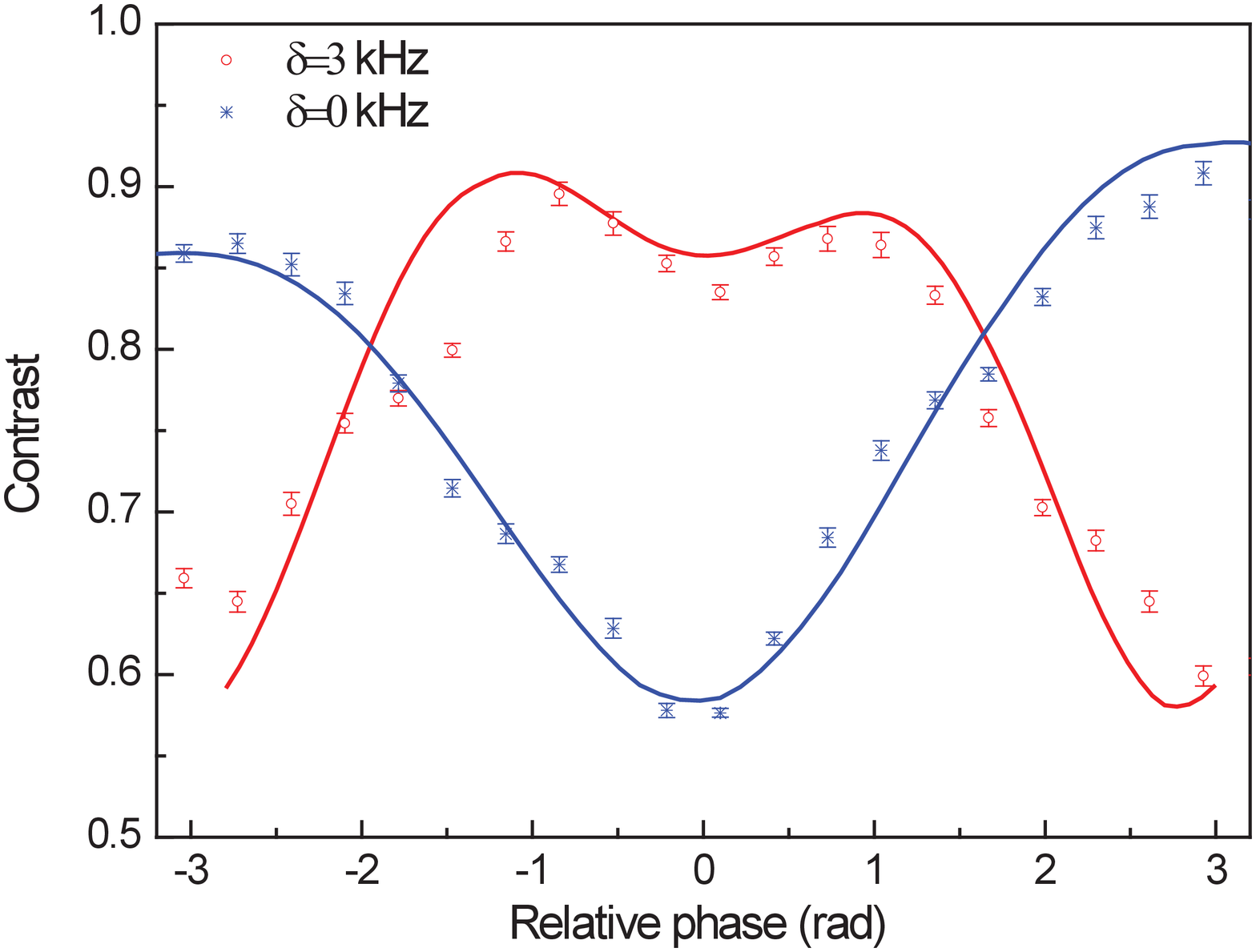}
	\caption{Contrast dependence on the phase of the middle $\pi$ Raman pulse for $T=160$ $\mu$s. The contrast is optimized by compensating the relative phases among multiple interference paths for $\delta=3$ kHz (red circles) and $\delta=0$ kHz (blue stars).}%
	\label{fig-5}%
\end{figure}

\vskip 5pt
\noindent

The contrast is investigated by modulating the relative phase of Raman pulses. When $\phi_{2}$ is modulated, the contrast is oscillating as shown in Fig.\textit{\ref{fig-5}}. Here, the contrasts are obtained by fitting the fringes for the different relative phase of the middle Raman pulse. For $\delta=3$ kHz, the phase shifts are induced by the two-photon detuning in the interference paths II and III. The contrast is minimum at $\phi_{2}=2.7$ rad because of the maximum phase competition between the M-Z ($\mathrm{I}$) and Ramsey ($\mathrm{II}$ and $\mathrm{III}$) interferometers. The maximum contrast is obtained at $\phi_{2}=-1.1$ rad, because the competition effect is suppressed by compensating the relative phases among the multiple interference paths. For $\delta=0$ kHz, the contrast varies as the phase of Raman lasers. The contrast is minimum at $\phi_{2}=-0.1$ rad, and is maximum at $\phi_{2}=3.0$ rad. For $\eta=82$ $\%$, $T=160$ $\mu$s and $\tau=30$ $\mu$s, the theoretical results are calculated according to Eq.(\ref{eq1}) (solid lines). The contrast strongly depends on the relative phases among three interference paths, which is verified by modulating the relative phase of Raman pulses. The contrast can be optimized by compensating the relative phases among the multiple interference paths.

\vskip 5pt
\noindent

\begin{figure}[htp]
	\centering
		\includegraphics[width=0.46\textwidth]{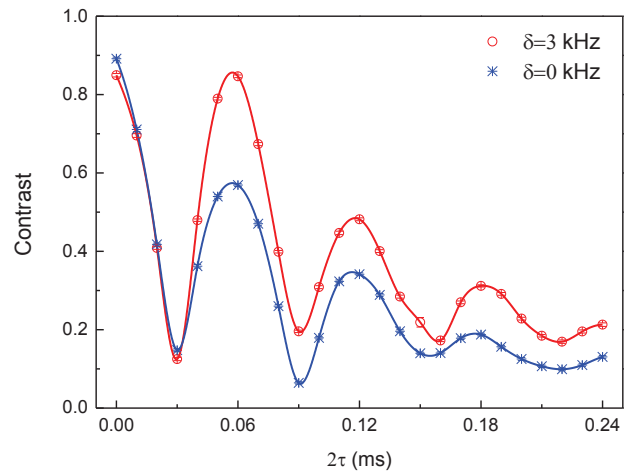}
	\caption{ Contrast dependence on the Raman durations for $T=160$ $\mu$s. The contrast oscillates with the same period for $\delta=0$ kHz (blue stars) and $\delta=3$ kHz (red circles).}%
	\label{fig-4}%
\end{figure}

\vskip 5pt
\noindent

When the interrogation time is fixed to $T=160$ $\mu$s in the M-Z interferometer, the contrast varies as the duration of the middle Raman pulse as shown Fig.\textit{\ref{fig-4}}.  The periodic oscillations are observed for $\delta=3$ kHz (red circles) and $\delta=0$ kHz (blue stars). The oscillation period is $60$ $\mu$s, which corresponding to the duration of the $\pi$ Raman pulse. For $\delta=3$ kHz, the M-Z interferometer has the almost same phase with the Ramsey interferometer ($\mathrm{II}$), but it has a phase difference with the Ramsey interferometer ($\mathrm{III}$). The competition of the M-Z interferometer is mainly from the Ramsey interferometer ($\mathrm{III}$). At $2\tau=0$ $\mu$s, the Ramsey interferometer ($\mathrm{III}$) only forms with the Raman pulse sequence ($\pi/2-\pi/2$). At $2\tau=30$ $\mu$s, the contrast decreases due to the strong competition between the M-Z interferometer ($\mathrm{I}$) and Ramsey interferometer ($\mathrm{III}$). For $\delta=0$ kHz, the M-Z interferometer competes with two Ramsey interferometers ($\mathrm{II}$ and $\mathrm{III}$). This leads that the contrast for $\delta=0$ kHz is lower than that for $\delta=3$ kHz. Due to the spontaneous emission, the contrasts decrease as the duration increased. By modulating the duration of Raman pulse, we confirm that the competition among multiple interference paths is mainly caused by the population transfer efficiency of stimulated Raman transition. The solid lines are the fitted results.

\section{conclusion}

In summary, we investigated the competition effect among the multiple quantum paths in the Raman-type M-Z interferometer. The competition of the multiple interference paths was theoretically and experimentally investigated by measuring the contrast of interference fringe. The contrast varies periodically as the interrogation time, duration and phase of Raman pulse. By modulating the duration of Raman pulse, we confirmed that the multiple quantum paths form because of the imperfect population transfer efficiency in stimulated Raman transitions. The contrast is influenced by the relative phases of multiple paths.
This work provides a method to investigate the competition effect caused by the low population transfer efficiency in stimulated Raman transitions. The phase competition can be suppressed by choosing the interrogation time and the parameters of Raman lasers, or may be avoided by improving the population transfer efficiency with stimulated Raman adiabatic techniques \cite{Kotru2014a,Kotru2015a}.

\vskip 5pt
\noindent

Although the phase competition is studied in the M-Z interferometer, this phenomenon does exist in multi-pulse interferometers due to the closed interference paths. For the co-propagating Raman beams, the competition effect should be considered in the spin echo sequence and in the Carr-Purcell-Meiboom-Gill sequence. For the counter-propagating Raman beams, the multiple interference loops form in the multi-pulse interferometer. The competition effect should also be taken into account in the multiple-pulse interferometer.  As an example, to improve the interference area, the composite-Raman-pulse sequence was used for designing multi-pulse interferometers \cite{Berg2015a}. The contrast could be optimized by modulating the relative phases of multiple interference paths. If the relative phase of composite pulses and the dark time between two consecutive composite pulses are reasonably chosen, the phase competitions among multiple interference paths could be suppressed, which is useful for designing multi-pulse interferometers.

\vskip 5pt
\noindent

We acknowledge the financial support from the National Key Research and Development Program of China under Grant No. 2016YFA0302002, the National Natural Science Foundation of China under Grant No. 11674362, No. 91536221, and No. 91736311, the Strategic Priority Research Program of Chinese Academy of Sciences under Grant No. XDB21010100, the Outstanding Youth Foundation of Hubei Province of China under Grant No. 2018CFA082, and the Youth Innovation Promotion Association of Chinese Academy of Sciences.

\bibliography{basename of .bib file}

\begin{thebibliography}{99}
\bibitem {ref2} Each data point is obtained by fitting one interference fringe. After the fringe is fitted by $y=y_{0}+ A \sin[\pi(x-x_{c})/\omega]$, $A$, $y_{0}$, $\delta_{A}$ and $\delta_{y_{0}}$ are obtained from the fitted results. The contrast is obtained by $C=A/y_{0}$, and the fitted error is given by $\delta=\sqrt{\delta_{A}^{2}+\delta_{y_{0}}^{2}}$. The data points and errors are given with the same method in Figs.\textit{\ref{fig-2}}-\textit{\ref{fig-5}}.
\end{thebibliography}

\end{document}